\documentclass[]{pasj01}

\Received{}
\Accepted{}
 
\usepackage[T1]{fontenc} 
\usepackage[switch,mathlines]{lineno}
 
\begin{document}

\title{
Anisotropic Thermal Conduction as a Driver of Jet Collimation and Magnetic Field Amplification on Cold Fronts
}

\author{Nana \textsc{Matsuno}\altaffilmark{1,2}}%
\altaffiltext{1}{Astronomical Science Program, Graduate Institute for Advanced Studies, SOKENDAI, 2-21-1 Osawa, Mitaka, Tokyo 181-8588, Japan}
\altaffiltext{2}{Division of Science, National Astronomical Observatory of Japan, 2-21-1 Osawa, Mitaka, Tokyo 181-8588, Japan}
\email{nana.matsuno@grad.nao.ac.jp}

\author{Takaaki \textsc{Yokoyama},\altaffilmark{3}}
\altaffiltext{3}{Astronomical Observatory, Kyoto University, Sakyo-ku, Kyoto, 606-8502, Japan}

\author{Mami \textsc{Machida}\altaffilmark{2,1}}



\maketitle


\begin{abstract}
Galaxy clusters contain a hot, diffuse, and weakly magnetized plasma known as the intracluster medium (ICM).
In this environment, how thermal conduction influences plasma dynamics and the conditions under which it operates efficiently remain open questions in cluster physics.
Systems in which active galactic nuclei (AGN) jets interact with cold fronts produced by cluster mergers provide a unique setting to examine the interplay between conduction, jet dynamics, and ordered magnetic fields.
To interpret the detailed structures revealed by recent observations, it is therefore important, as a first theoretical step, to quantify how thermal conduction modifies AGN jet morphology and the surrounding magnetic-field configuration.
We perform two-dimensional magnetohydrodynamic (MHD) simulations of an AGN jet in an ICM environment, incorporating anisotropic thermal conduction with varying efficiency. 
The simulations show that thermal conduction transports heat from the jet head backward along magnetic field lines into the inner cocoon.
This process increases the inner cocoon pressure, enhancing jet collimation by a factor of $\sim 4$ compared to models without conduction. 
This stronger collimation stretches the magnetic fields along the cold-front surface, resulting in a maximum field strength up to a factor of $\sim 1.5$ larger. 
Jet collimation increases as the conduction efficiency increases, which is interpreted as a \textit{conductive collimation mechanism}.
These results suggest that anisotropic thermal conduction can operate effectively on jet scales in galaxy clusters, and that accounting for conduction may be important when interpreting jet morphology and magnetic field structure in merging cluster environments.
\end{abstract}


\section{Introduction}

Galaxy clusters contain a hot, diffuse, and weakly collisional intracluster medium (ICM) in which thermal conduction has long been recognized as a potentially important energy transport process (e.g., \cite{Sarazin88}; \cite{Narayan01}; \cite{Voit05}).
 Classical estimates based on the Spitzer conductivity~\citep{Spitzer62} indicate that the conductive timescale in cluster cores is of order $10^8$~yr, shorter than the typical dynamical timescale of the system~\citep{Sarazin88}. 
 On the other hand, observational studies indicate that the actual thermal conduction in the radial direction within clusters must be strongly suppressed relative to the Spitzer value. 
Specifically, analyses of temperature fluctuations and radial profiles suggest that the global effective conductivity of the ICM is reduced by factors of up to several tens (\cite{Markevitch03}; \cite{Fang18}).
 In a magnetized plasma, heat is transported primarily along magnetic field lines~\citep{Braginskii65}, hence the effective conductivity is expected to depend sensitively on the local magnetic-field geometry~\citep{Narayan01}.
Thus, the conditions under which thermal conduction operates efficiently in the ICM, and the ways in which it influences the dynamics of the plasma, remain open questions in the physics of galaxy clusters.

To illustrate how thermal conduction behaves under different physical conditions in the ICM, we focus in this study on the properties of cold fronts and active galactic nuclei (AGN) jets.
First, cold fronts provide a typical example in which thermal conduction is strongly suppressed~(see the review of \cite{Markevitch07}).
They are observed as sharp contact discontinuities in gas density and temperature~\citep{Henry96}, typically arising when a cool, dense sub-cluster core survives its passage through a hotter ambient cluster or from sloshing motions in a disturbed core~(e.g., \cite{Markevitch07}).
This motion stretches and compresses the ambient magnetic field ahead of the core, forming a strongly ordered magnetic layer aligned with the cold-front surface~(magnetic draping; e.g., \cite{Asai04}). 
Magnetohydrodynamic (MHD) simulations have confirmed that such draping layers on the surface of cold fronts can stabilize the front and inhibit cross-field thermal transport~(Asai et al.\ \yearcite{Asai04}, \yearcite{Asai05}, \yearcite{Asai07}).
In contrast, the role of thermal conduction in AGN jets propagating through the ICM is much less well understood. 
Magnetic draping is expected to occur at contact discontinuities associated with astrophysical jets, where ambient magnetic fields are compressed and stretched along the jet boundary (e.g., \cite{Lyutikov06}; \cite{Dursi08}).
Such field amplification can produce anisotropic transport conditions.
Theoretical studies also suggest that shocks within AGN jets can heat localized regions and create steep temperature gradients~(e.g., \cite{Norman82}), in which thermal conduction may influence the internal jet structure. 
In addition, MHD simulations on accretion-disk scales have reported that thermal conduction can enhance jet collimation and acceleration~(\cite{Rezgui22}; \cite{Rezgui25}). 
However, on galaxy-cluster scales, although the impact of AGN activity on the surrounding ICM has been investigated~(e.g., \cite{Parrish09}; \cite{Beckmann22}), how thermal conduction acts within the jet itself and modifies its internal structure remains poorly understood.

The behavior of thermal conduction in cold fronts and AGN jets can be examined in greater detail by studying galaxy clusters in which these two structures interact directly. 
Recent multi-wavelength observations have revealed clear examples of jets that come into contact with, and are distorted by cold fronts~(Chibueze, Sakemi, Ohmura et al. 2021, hereafter C21; \cite{Giacintucci22}). 
C21 reported that the jet emitted from the radio galaxy MRC~0600--399 in Abell~3376 shows a pronounced $\sim 90^\circ$ change in jet direction at the location of the cold front, and then propagates in a highly collimated manner over a distance of $\sim 100~{\rm kpc}$. 
They interpreted this morphology as the jet being bent by magnetic fields draped along the cold-front surface, with cosmic rays subsequently propagating along those ordered field lines. 
Furthermore, a polarimetry of MRC~0600--399 by \citet{Sakemi25} showed that the radio emission at the tip of the jet is tightly associated with the magnetic field on the cold-front surface, lending additional support to the C21 interpretation. 
These results suggest that the intersection between the jet and the cold front offers a natural laboratory in which to test not only the magnetic-field structure but also the role of thermal conduction by confronting multi-wavelength observations with analytical and numerical models. 
Despite these observational advances, the role of thermal conduction inside the jet and at its interface with the cold front remains unclear.
To interpret the detailed structures revealed by observations, it is therefore important, as a first step on the theoretical and simulation side, to quantify how thermal conduction alters the jet morphology and the surrounding magnetic-field configuration.

In this study, we perform two-dimensional MHD simulations that include thermal conduction and model a jet impacting an ICM environment designed to mimic the magnetic-field configuration expected at a cold-front surface. 
Our aim is to investigate how thermal conduction alters the jet morphology and influences the jet–cold-front interaction, and to identify the underlying physical processes. 
A detailed comparison with actual observations is beyond the scope of this work and is left for future studies. 
We also examine the dependence of these effects on the conduction efficiency by systematically varying its value.
Our work extends the numerical simulations of C21 by incorporating thermal conduction.

Section~2 describes the numerical simulation setup, including the initial conditions and the implementation of anisotropic thermal conduction.
In section~3, we present the simulation results, focusing on the differences in jet morphology and magnetic field evolution with and without thermal conduction,  as well as the dependence on the conduction efficiency. 
Section~4 discusses the implications of our findings for AGN jets and cold-front physics in galaxy clusters and outlines the limitations of our simulations.  
Finally, section~5 summarizes our conclusions and prospects for future work.

\section{Numerical Methods}
We performed MHD simulations including thermal conduction to study the time evolution of a jet propagation through the ICM.
\subsection{Basic Equations}
We solved the MHD equations including the thermal condition: 
\begin{equation}
\frac{\partial \rho}{\partial t}
+ \nabla \cdot \left( \rho \boldsymbol{v} \right) = 0,
\end{equation}
\begin{equation}
\frac{\partial (\rho \boldsymbol{v})}{\partial t}
+ \nabla \cdot \left[
\rho \boldsymbol{v}\boldsymbol{v}
+ \left( p + \frac{\boldsymbol{B}^2}{8\pi} \right)\mathcal{I}
- \frac{\boldsymbol{B}\boldsymbol{B}}{4\pi}
\right] = 0,
\end{equation}
\begin{equation}
\frac{\partial\boldsymbol{B}}{\partial t}=-c \nabla\times{\boldsymbol{E}},
\end{equation}
\begin{equation}
\begin{array}{l}
\frac{\partial}{\partial t}\left(e_{\rm int} + \frac{1}{2}\rho\boldsymbol{v}^2 + \frac{\boldsymbol{B}^2}{8\pi}\right) + \\
\nabla\cdot\left[\left(e_{\rm int}+p \right)\boldsymbol{v} + \frac{\rho \boldsymbol{v}^2}{2}\boldsymbol{v} + \frac{c}{4\pi}\boldsymbol{E}\times\boldsymbol{B} + \boldsymbol{F}_{\rm TC}\right]=0,
\end{array}
\label{eq:energy-equation}
\end{equation}
\begin{equation}
\boldsymbol{E} = -\frac{1}{c}\boldsymbol{v} \nabla\times \boldsymbol{B},
\end{equation}
\begin{equation}
    e_{\rm int } = \frac{p}{\gamma -1},
\end{equation}
where $t$, $\rho$, $\boldsymbol{v}$, $p$, $\boldsymbol{B}$, $\boldsymbol{E}$, $c$ and $\gamma = 5/3$ are time, density, velocity, pressure, magnetic fields, electric fields, the light speed, and the specific heat ratio respectively. 
$\mathcal{I}$ represents the identity tensor. 
We ignore radiative cooling, resistivity and viscosity in this paper. 
We adopt the ideal gas law and $T$ is computed by the ideal gas equation of state, 
\begin{equation}
    T = \frac{m_{\rm p} \mu}{k_{\rm B}} \frac{p}{\rho},
\end{equation}
where $m_{\rm p}$ is the atomic mass unit, $\mu = 0.60$ is the mean molecular weight, and $k_{\rm B}$ is the Boltzmann constant.

The thermal conduction fluxes ($\boldsymbol{F}_{\rm TC}$), in the energy equation, are evaluated by two types of heat fluxes, the classical flux and the saturation flux. Following \citet{Spitzer62}, the classical flux for thermal conduction can be written as 
\begin{equation}
    \boldsymbol{F}_{\rm class} = - \kappa T^{5/2}\boldsymbol{b} \left(\boldsymbol{b}\cdot \nabla T\right),
\end{equation}
 where $\kappa = 4.6 \times 10^{-7}~{\rm erg~cm^{-1}~s^{-1}~K^{-7/2}}$ and $\boldsymbol{b}$ are the coefficients of thermal conduction and the unit vector of the magnetic fields, respectably. 
An extension of the equations in \citet{Cowie77} that preserves full vector consistency shows that the saturated flux for thermal conduction can be written as 
\begin{equation}
    \boldsymbol{F}_{\rm sat} = - 5\phi \rho c_{\rm s}^{3} ~{\rm sgn}\left( \boldsymbol{b}\cdot \nabla T \right)\boldsymbol{b},
\end{equation}
where $c_{\rm s}=\sqrt{\gamma p/\rho}$ is the sound speed. 
Following \citet{Balbus82}, we adopt $\phi=0.3$, which represents a commonly used order-unity coefficient describing the efficiency of saturated electron heat flux in weakly collisional plasmas.
The total thermal conduction flux can be written as (see also~\citet{Meyer12})
\begin{equation}
    \boldsymbol{F}_{\rm TC} =  f \left\{ G\left(\frac{|\boldsymbol{F}_{\rm class}|}{|\boldsymbol{F}_{\rm sat}|}\right)\boldsymbol{F}_{\rm class} + \left[ 1- G\left(\frac{|\boldsymbol{F}_{\rm class}|} {|\boldsymbol{F}_{\rm sat}|}\right) \right]\boldsymbol{F}_{\rm sat} \right\},
\label{eq:F_TC}
\end{equation}
where the flux limiter, $G(a)$, is given by
\begin{equation}
    G(a) = \max\{ 0, \min\left[1, 1-0.5a\right] \}.
\end{equation}
$f$ is the conduction efficiency depending on the models (details are written in subsection \ref{sec:model_TC}).

\subsection{Numerical Methods and Boundary Conditions}
We solved the MHD equations using the Harten-Lax-van Leer (HLL) approximate Riemann solver \citep{Einfeldt91}. 
Reconstruction adopts Monotone Upstream–centered Schemes for Conservation Laws (MUSCL;~\cite{vanLeer79}) to achieve second-order accuracy in space. 
For the time integration of the MHD equations, we employ the second-order midpoint method.
Additionally, the hyperbolic divergence cleaning method is adopted for the magnetic field~\citep{Dedner02}. 
To calculate the thermal conduction term in the energy equation (\ref{eq:energy-equation}), we adopt the Super-Time-Stepping~(STS) method (Meyer et~al.\ \yearcite{Meyer12}, \yearcite{Meyer14}). 
It solves the thermal conduction with second-order temporal and spatial accuracy. 

We adopted two-dimensional Cartesian coordinates ($x, y$) by assuming translational symmetry in the $z$-direction and only $x$ and $y$ components for velocity and magnetic fields. 
The domain of the simulation box and the number of grid points are $\left(-156\ {\rm kpc}\le x \le 156\ \mathrm{kpc}, 0\ \mathrm{kpc} \le y \le 196~\mathrm{kpc}\right)$, and $\left(N_x, N_y\right)$ $= \left(1248,784\right)$. 
So the width of each uniform grid is $\Delta x = \Delta y = 0.25\ \mathrm{kpc}$. 
Here, the $y$-direction is aligned with the jet axis. 
We impose Neumann-type boundary conditions (i.e., zero-gradient conditions) on all sides of the computational domain, meaning that the normal derivatives of all physical variables vanish at the domain boundaries. 

\subsection{Initial Model}
At the initial state, we assumed that the jet propagates into the homogeneous static ICM and hits the arch-like strong fields, whose inner boundary is considered to correspond to the cold front observed in the galaxy cluster.
Based on~C21, we set the magnetic fields: 
\begin{equation}
    B_{x} = -\frac{x}{r} B_\theta,~~B_{y} = \frac{y}{r} B_\theta,~~B_{z} = 0,
\end{equation}
where $B_\theta$ is the circumference magnetic field in polar coordinates,
\begin{equation}
  B_\theta =
  \left\{
  \begin{array}{@{}ll@{}}
    B_{\mathrm{ICM}} + B_{1}\sin\!\left[\frac{(r-r_{\rm s})\pi}{w}\right]
    & (r_{\rm s} \leq r \leq r_{\rm s}+w), \\
    B_{\mathrm{ICM}}
    & (r < r_{\rm s},~r_{\rm s}+w < r).
  \end{array}
  \right.
\end{equation}
where $r=\sqrt{x^2 + y^2},~\theta =\arctan\left( y/x\right)$ is the distance from the origin and the angle from the $x$-axis, $B_{\mathrm{ICM}} = 0.7~\mu{\rm G}$ is the magnetic fields strength in ICM, $B_{1} = 18~\mu{\rm G}$ is the peak fields of the magnetic arch at $r_{\rm  s}=110~\mathrm{kpc}$, and $w=30~\mathrm{kpc}$ is the width of the arch, respectively. 

ICM pressure profile is determined by the equilibrium condition in the polar coordinates system,
\begin{equation}
p(r) = p(r=0) - \frac{B^{2}_{\theta}(r)}{8\pi}-\int^{r}_{0}\frac{B^{2}_{\theta}(r')}{4\pi r'}dr', 
\label{eq:InitialPressure}
\end{equation}
where $p(r=0) = 3.1\times 10^{-11}~{\rm erg~cm^{-3}}$. 
We assume the ICM has homogeneous density $\rho = 8.0 \times 10^{-27}~{\rm g~cm^{-3}}$. 

The jet is injected from the boundary at $(x_{\rm jet}, y_{\rm jet}) = (60~{\rm kpc}, 0~{\rm kpc})$ over a finite width in the $x$-direction.
Specifically, the injection region is defined by $|x - x_{\rm jet}| < d_{\rm jet}$, where $d_{\rm jet} = 3~{\rm kpc}$.
The density, pressure, and velocities of the injected flow are $\rho_{\rm jet} = 8.0\times 10^{-27}~{\rm g~cm^{-3}}, p_{\rm jet} = 3.1\times 10^{-11}~{\rm erg~cm^{-3}}$, and $v_{y,{\rm jet}} = 5.0 \sqrt{\gamma p_{\rm jet}/\rho_{\rm jet}} \simeq 4000~{\rm km~s^{-1}},~v_{x,{\rm jet}}=0~{\rm cm~s^{-1}}$ respectively. The jets do not have any magnetic fields. 
To avoid artificially enhanced thermal transport along the magnetic field lines from the injection region, we set the thermal conduction efficiency to zero within a radius of $10~{\rm kpc}$ from the jet injection point.

\subsection{Modeling Thermal Conduction in the ICM}
\label{sec:model_TC}
To systematically assess the effects of thermal conduction, we carried out simulations with varying values of $f$ in equation~(\ref{eq:F_TC}), ranging from $0.0$ (no conduction) to $1.0$ (full conduction):
$f = { 0.0,~0.005, ~0.01, ~0.05, ~0.1,~0.3, ~0.5, ~1.0}.$
In particular, the models with $f = 0.0$ and $f = 1.0$ are referred to as "w/o~TC" (without thermal conduction) and "w/~TC" (with thermal conduction), respectively.
Values of $f < 1.0$ are intended to model the suppression of conduction caused by tangled magnetic fields or turbulence in the ICM.

\section{Numerical Results}

\subsection{Time Evolution Without Thermal Conduction}

\begin{figure*}[ht]
    \centering
    \includegraphics[width=\linewidth]{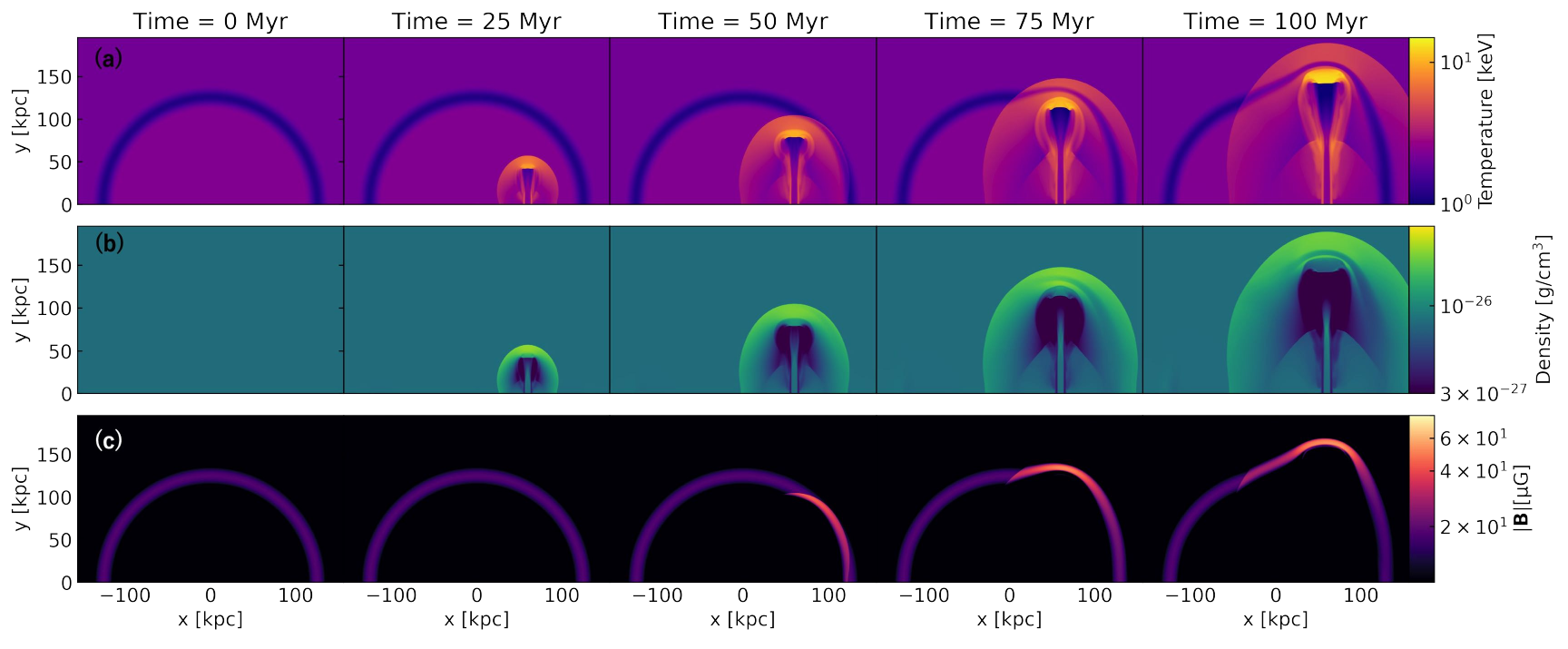}
    \caption{
        Temporal evolution of (a) temperature, (b) density,and (c) magnetic field strength in the model w/o~TC. Snapshots are shown at $t = 0$, 25, 50, 75, and 100~${\rm Myr}$. A corresponding animation is provided as an online supplementary E-movie~1.
    {Alt text: Time evolution of a two dimensional magnetohydrodynamic jet simulation without thermal conduction. Rows show temperature in kiloelectronvolts, gas density in grams per cubic centimeter, and magnetic field strength in microgauss. Spatial axes are in kiloparsecs.}
    }
    \label{fig:without_tc}
\end{figure*}

Figure~\ref{fig:without_tc} shows the time evolution of temperature, density, pressure, and magnetic field strength in the model w/o~TC.
The AGN jet propagates upward while maintaining an axisymmetric morphology and interacting with the magnetic field at the cold front.
As the jet advances, it stretches and amplifies the magnetic field near the cold fronts.
These results confirm that the interaction between the jet and the cold front can lead to significant magnetic field amplification, even in the absence of thermal conduction.
This result is consistent with the findings of C21.

\subsection{Time Evolution With Thermal Conduction}

\begin{figure*}[ht]
    \centering
    \includegraphics[width=\linewidth]{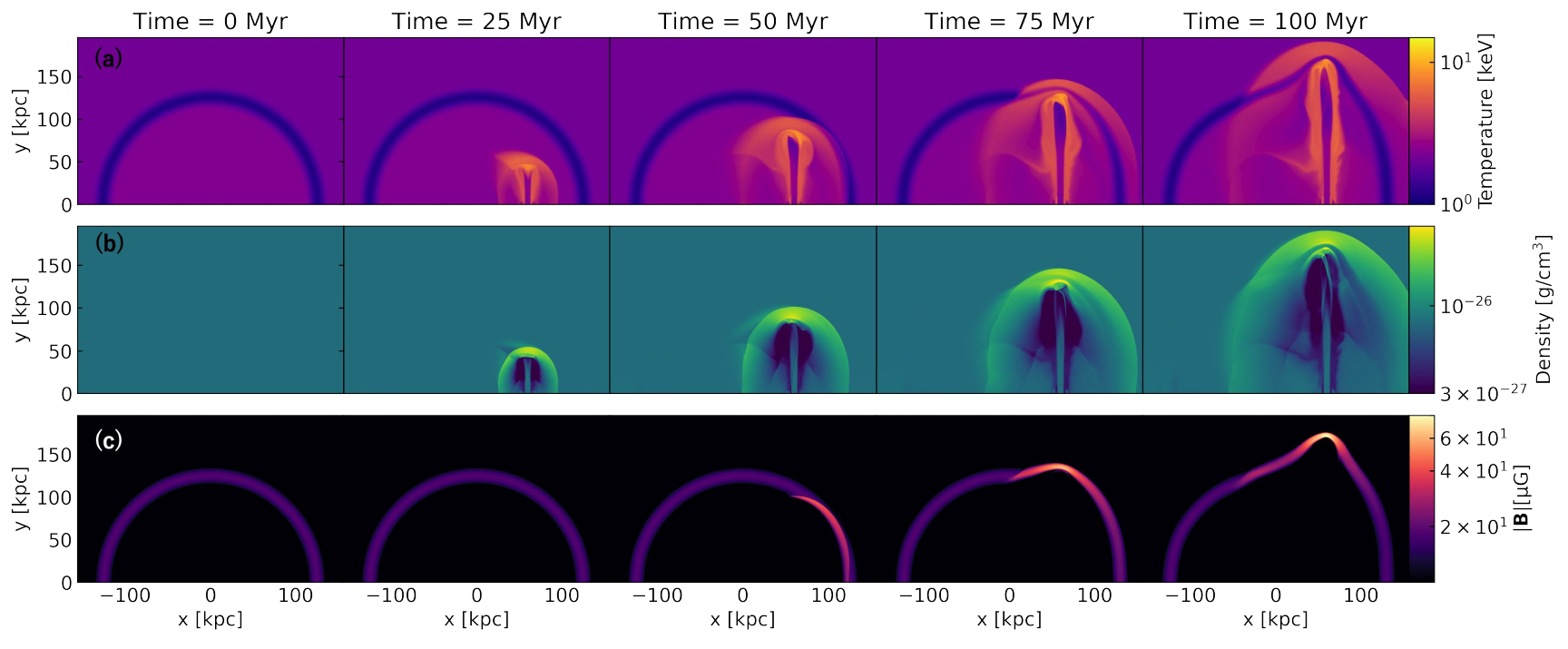}
    \caption{
        Temporal evolution of (a) temperature, (b) density, and (c) magnetic field strength in the model w/~TC. Snapshots are shown at $t = 0$, 25, 50, 75, and 100 ${\rm Myr}$. A corresponding animation is provided as an online supplementary E-movie~2.
        {Alt text: Time evolution of a two dimensional magnetohydrodynamic jet simulation with thermal conduction. Rows show temperature in kiloelectronvolts, gas density in grams per cubic centimeter, and magnetic field strength in microgauss. Spatial axes are in kiloparsecs.}
    }
    \label{fig:with_tc}
\end{figure*}

Figure~\ref{fig:with_tc} shows the results for the model w/~TC.
As in the model without conduction, the jet propagates upward and interacts with the cold front, leading to magnetic field amplification.
However, notable differences emerge in the jet morphology and the spatial distribution of the magnetic field.
The jet structure becomes narrower and more asymmetric in terms of the jet axis over time~(see subsection~\ref{sec:result_jet}), and the amplified magnetic field near the cold front exhibits distinct spatial features~(see subsection~\ref{sec:result_mag}).
These differences highlight the significant influence of thermal conduction on jet dynamics and magnetic field evolution.

\subsection{Jet Structure}
\label{sec:result_jet}
\begin{figure}
    \centering
    \includegraphics[width=\linewidth]{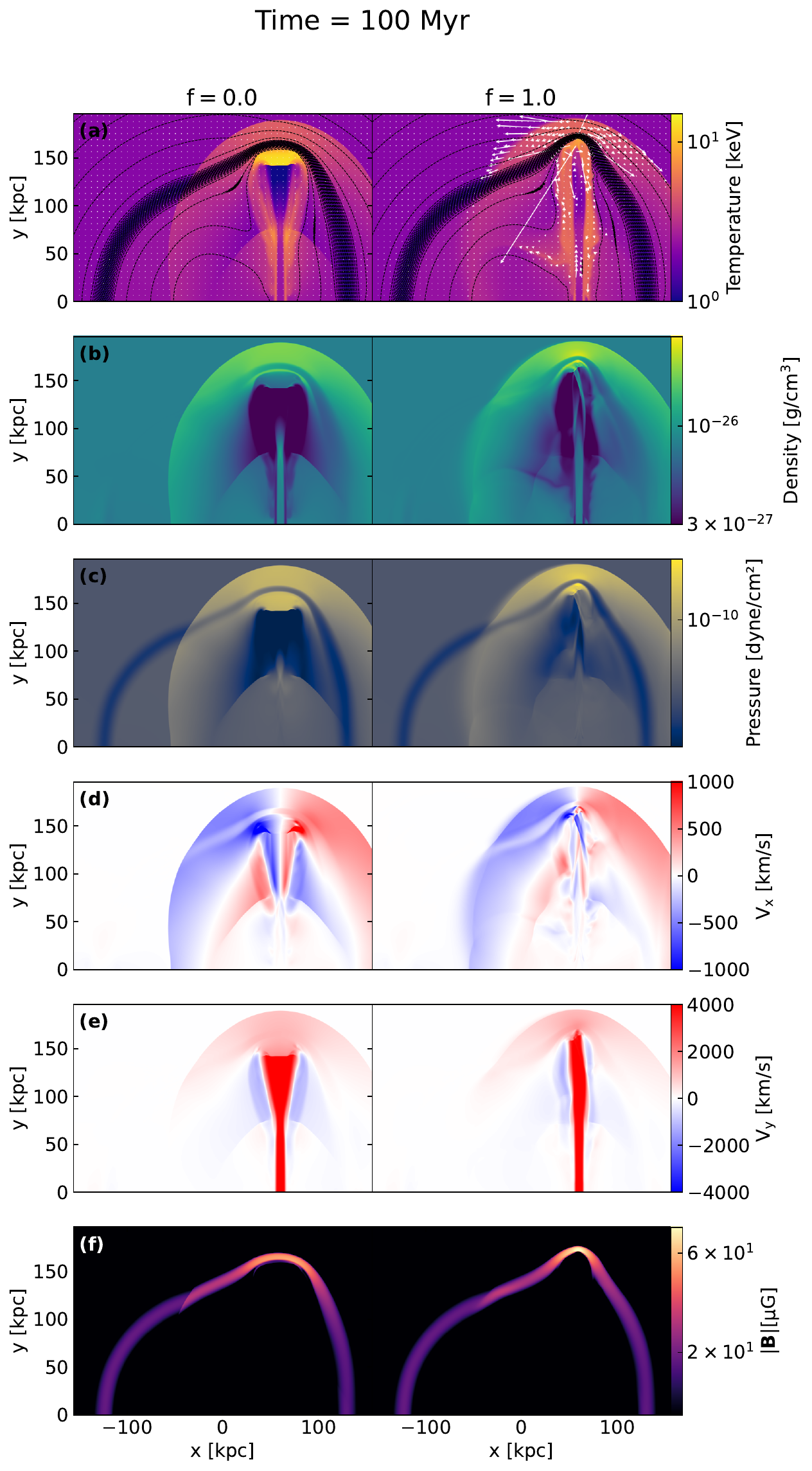}
    \caption{Comparison between the models w/o~TC (left panels) and w/~TC (right panels) at $t = 100~{\rm Myr}$. 
  Panels (a)–(f) show the two-dimensional distributions of (a) temperature, (b) pressure, (c) density, (d) $x$-component of velocity, (e) $y$-component of velocity, and (f) magnetic field strength, respectively.
  In panel (a), magnetic field lines (black dashed lines) and thermal-conduction flux vectors (white arrows) are overplotted.
    {Alt text: Comparison of two dimensional magnetohydrodynamic jet simulations without and with thermal conduction at one hundred megayears. Left panels show the model without thermal conduction and right panels show the model with thermal conduction. Rows present temperature in kiloelectronvolts, pressure in dyne per square centimeter, gas density in grams per cubic centimeter, x component of velocity in kilometers per second, y component of velocity in kilometers per second, and magnetic field strength in microgauss. Spatial axes are in kiloparsecs.}
  }
    \label{fig:comparison_f0_f1}
\end{figure}

\begin{figure*}
    \centering
    \includegraphics[width=\linewidth]{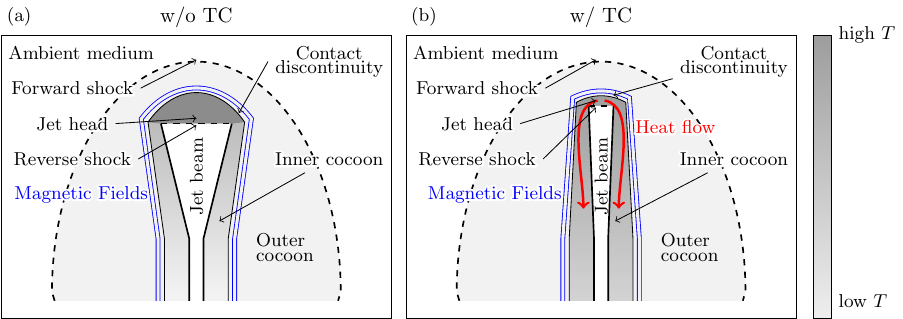}
    \caption{Schematic comparison of the jet morphology for (a) w/o~TC and (b) w/~TC.
    {Alt text: Schematic comparison of jet morphology in two dimensional magnetohydrodynamic models without and with thermal conduction. The left panel shows the model without thermal conduction and the right panel shows the model with thermal conduction. Each panel illustrates the ambient medium, jet beam, jet head, forward shock, reverse shock, contact discontinuity, inner cocoon, and outer cocoon. Blue lines show the magnetic fields, and red vectors show heat flow.}
    }
    \label{fig:schematic_comparison_of_jet}
\end{figure*}

In this section, we compare the results of the w/o~TC and w/~TC models, focusing on the changes in jet structure at $100~{\rm Myr}$. 
The jet consists of a collimated beam of supersonic plasma that propagates through the ambient ICM, extending from the injection point at $x=60~{\rm kpc}, y=0~{\rm kpc}$ to $y\approx140~{\rm kpc}$ (figure~\ref{fig:comparison_f0_f1}a; See also the schematic illustration in figure~\ref{fig:schematic_comparison_of_jet}). At the leading edge of the beam, its interaction with the ambient ICM produces a forward shock that compresses the ICM ahead~(around $x\approx60~{\rm kpc}, y\approx170~{\rm kpc}$).
Just behind this, the jet flow is decelerated by a reverse shock~(around $x\approx60~{\rm kpc}, y\approx140~{\rm kpc}$), and the region between these two shocks, containing shocked jet plasma and shocked ambient ICM, constitutes the jet head~($x\approx60~{\rm kpc}, y=140$--$170~{\rm kpc}$).
Around the jet beam, a low-density inner cocoon forms as shocked jet material expands laterally into the surrounding shocked ICM, refereed to as the outer cocoon.

\begin{figure}
    \centering
    \includegraphics[width=\linewidth]{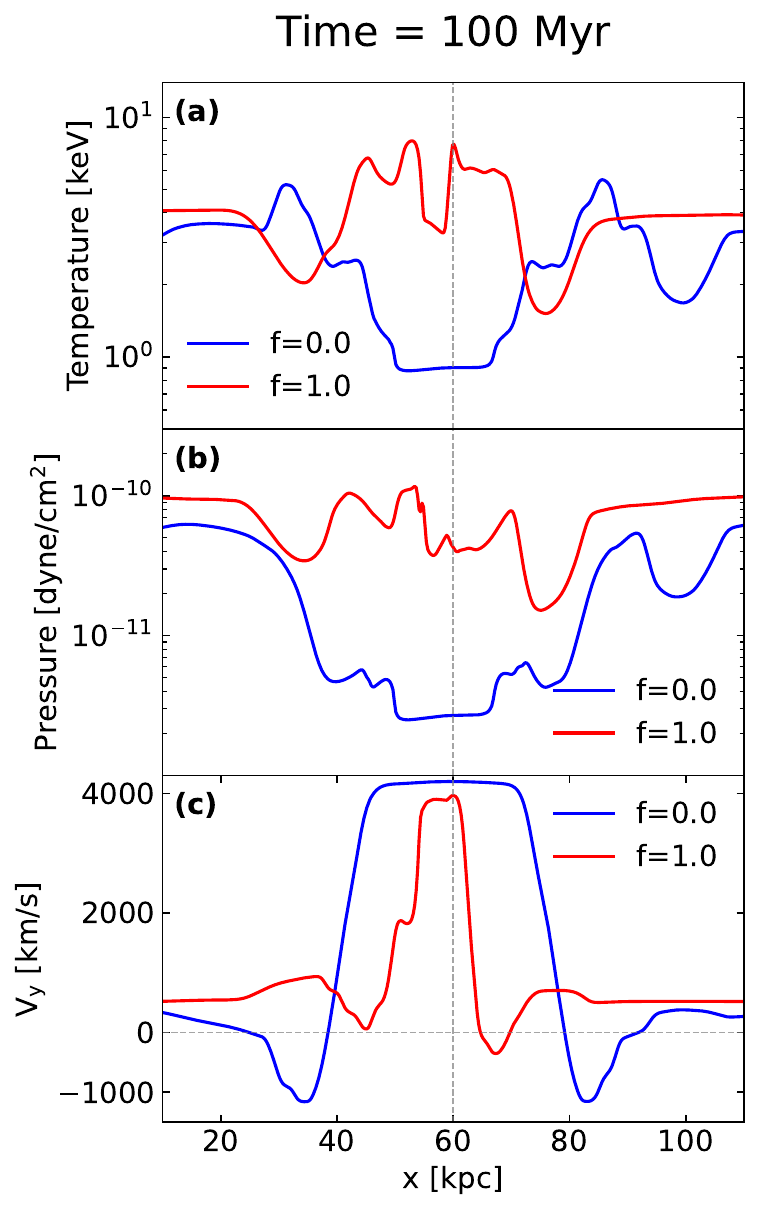}
    \caption{One-dimensional profiles of (a) temperature, (b) pressure and (c) $y$-component of velocity along the $x$-axis at $y= 125~{\rm kpc}$ and $t =100~{\rm Myr}$.
    {Alt text: One dimensional profiles at one hundred megayears along the x direction at a fixed y position. Panels show temperature in kiloelectronvolts, pressure in dyne per square centimeter, and y component of velocity in kilometers per second for models without and with thermal conduction. The horizontal axis represents distance in kiloparsecs.}
    }
    \label{fig:y-1d_125kpc}
\end{figure}

A key difference between the two models arises from the presence of anisotropic thermal transport in the w/~TC case, which redistributes thermal energy along the magnetic field lines.
As the jet propagates, magnetic draping leads to the accumulation of magnetic field lines along the contact discontinuity between the inner and outer cocoons (figure~\ref{fig:comparison_f0_f1}a), consistent with the scenario proposed by \citet{Lyutikov06}.
This configuration strongly suppresses thermal exchange across the discontinuity and confines heat transport primarily along the jet beam lines.
Consequently, thermal conduction transports heat from the jet head backward along the beam toward cooler regions, rather than allowing thermal energy to accumulate locally at the jet head~(figure \ref{fig:comparison_f0_f1}a).
The temperature in the inner cocoon, measured on a slice taken slightly upstream of the reverse shock --- at $y=130~{\rm kpc}$ for the w/o~TC model and $y=140~{\rm kpc}$ for the w/~TC model, corresponding to approximately $5~{\rm kpc}$ ahead of the reverse shock in each case --- therefore exhibits model-dependent spatial variations~(figure~\ref{fig:y-1d_125kpc}a).
In the w/o~TC model, the temperature remains relatively low at $x = 25$–$40$ and $80$–$95~{\rm kpc}$, while in the w/~TC model, pronounced temperature enhancements appear at $x = 40$–$50$ and $65$–$75~{\rm kpc}$.
Thermal redistribution leads to a substantial increase in the inner cocoon pressure, with the w/~TC model exhibiting values approximately four times higher than those in the w/o~TC model (figure~\ref{fig:comparison_f0_f1}a, \ref{fig:comparison_f0_f1}c, \ref{fig:y-1d_125kpc}a, \ref{fig:y-1d_125kpc}b).

The enhanced pressure within the inner cocoon in the w/~TC model has a direct impact on the jet morphology and kinematics. 
In this model, the jet beam becomes significantly more collimated compared to the w/o~TC case. 
Just upstream of the reverse shock, the jet-beam width decreases from $30~{\rm kpc}$ in the w/o~TC model to $7.8~{\rm kpc}$ in the w/~TC model, corresponding to a reduction by approximately a factor of four (figure~3e; figure~5-2c).
In addition to the geometric narrowing of the beam, the transverse velocity component within the jet beam is also reduced in the w/~TC model (figure~3d). 
As a consequence, the jet momentum is increasingly confined to the jet propagation direction, allowing the jet to advance more efficiently (figure~3e). 
This collimated propagation is accompanied by enhanced local pressure and density near the jet head in the w/~TC model, reflecting the modified dynamical state of the jet–cocoon system.

In summary, thermal conduction does not merely smooth temperature gradients, but instead redistributes heat backward along the magnetic field, thereby weakening lateral expansion while promoting forward propagation.
This single thermal-transport effect simultaneously produces the narrower cocoon and the enhanced jet penetration seen in the conductive model.

\subsection{Magnetic Field Amplification on Cold Fronts}
\label{sec:result_mag}

\begin{figure}
    \centering
    \includegraphics[width=\linewidth]{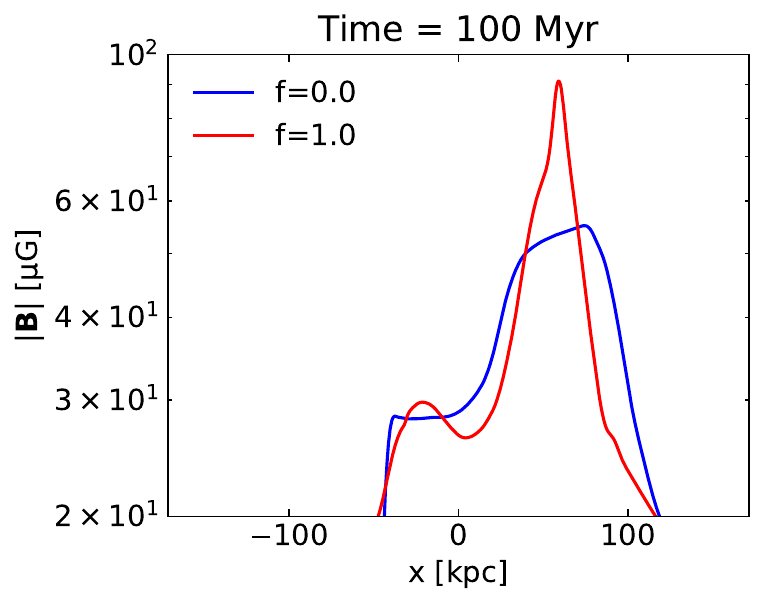}
    \caption{One-dimensional profile of the magnetic field strength at $t=100~\mathrm{Myr}$, obtained by extracting the maximum value along the $y$-direction at each $x$-coordinate.
    {Alt text: One dimensional profile of magnetic field strength at one hundred megayears as a function of distance along the x direction. The vertical axis shows magnetic field strength in microgauss and the horizontal axis shows distance in kiloparsecs. Profiles are shown for models without and with thermal conduction.}}
    \label{fig:1d_mag}
\end{figure}

Magnetic field amplification at cold-front surface around the jet head becomes more spatially localized when thermal conduction is included~(figure~\ref{fig:comparison_f0_f1}f).
In particular, thermal conduction suppresses lateral expansion of the shocked jet material, resulting in magnetic field lines being stretched preferentially along the jet axis rather than dispersed over a wider region.
As described in subsection~\ref{sec:result_jet}, thermal conduction transports heat backward along the jet axis from the jet head, reducing the local pressure peak and suppressing lateral expansion.
This results in a jet flow more tightly focused along the axis, which stretches magnetic field lines primarily in that direction.

This difference is also evident in the one-dimensional profile along the cold-front surface shown in figure~\ref{fig:1d_mag}.
The w/o~TC model exhibits magnetic field enhancement over a broad region ($x = 30$--$100~{\rm kpc}$), whereas the w/~TC model displays a sharper and more localized amplification centered at $x = 40$--$70~{\rm kpc}$, corresponding to the region where the jet approaches the cold front.
These results demonstrate that magnetic field amplification depends not only on shock compression but also on the flow geometry shaped by thermal conduction.
By promoting axial flow and limiting lateral expansion, thermal conduction leads to spatially confined magnetic field stretching around the jet head.

\subsection{Dependence on Conduction Efficiency}
\label{sec:result_f}

\begin{figure}
    \centering
    \includegraphics[width=\linewidth]{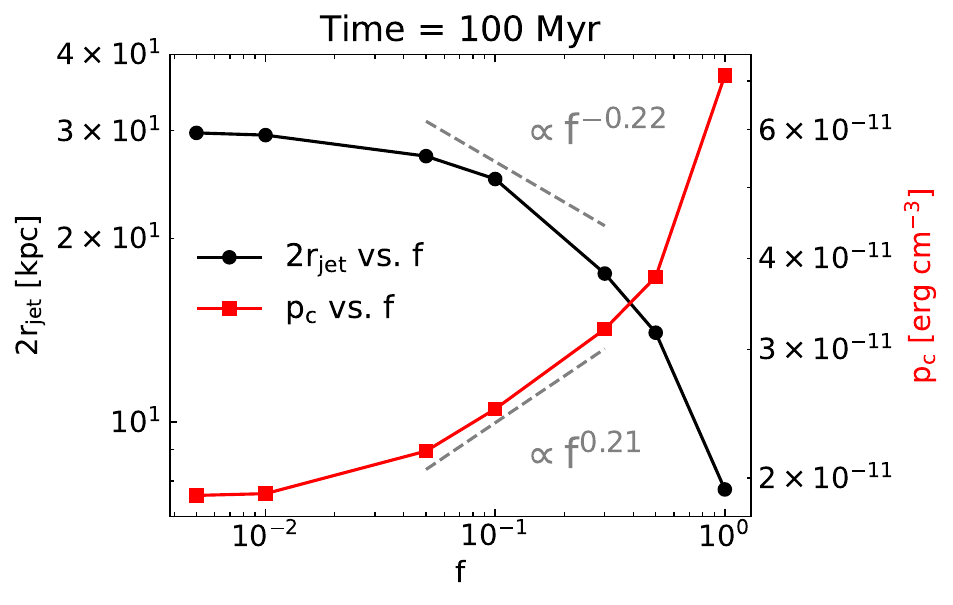}
    \caption{Jet beam width (2$r_{\rm jet}$) and pressure in the inner cocoon ($p_{\rm c}$) at $t=100~{\rm Myr}$ for models with different thermal conduction efficiency parameters $f$.
    {Alt text: Relationship between thermal conduction efficiency and jet properties at one hundred megayears. The plot shows jet beam width in kiloparsecs and pressure in the inner cocoon in dyne per cubic centimeter as functions of the thermal conduction efficiency parameter. Results are shown for multiple model values of the efficiency parameter.}}
    \label{fig:f_rjet_Pc}
\end{figure}

To examine how the efficiency of thermal conduction influences the system, we conduct a series of simulations with the conduction efficiency parameter $f$ varying from 0.0 to 1.0~(subsection~\ref{sec:model_TC}).
We then examine how this variation influences the pressure in the inner cocoon, $p_{\rm c}$, the jet width, $2r_{\rm jet}$, the maximum magnetic energy density on the cold-front surface, $e_{\rm mag}$, and the kinetic energy density at the jet head, $e_{\rm kin}$.

To quantify these quantities in a consistent manner, we first identify the characteristic regions used in our analysis. 
A schematic illustration of these regions is shown in Figure~\ref{fig:schematic_comparison_of_jet}.
The reverse shock along the jet beam is identified by averaging the jet-aligned velocity component $v_y$ within a 5~kpc wide strip centered on the jet axis and locating the position where the averaged $v_y$ drops below $3000~{\rm km~s^{-1}}$, corresponding to 75\% of the injection velocity. 
 A transverse slice located $5\ {\rm kpc}$ upstream of the reverse shock is used to define the jet beam and the inner cocoon, while the jet head is defined as the region located $5\ {\rm kpc}$ downstream of the reverse shock. 
 The interaction point between the jet and the cold front is defined as the location where the magnetic field strength is maximized near $x=60\ {\rm kpc}$.
 Specifically, for each position along the jet axis , we compute the magnetic field strength averaged over the range $55\ {\rm kpc}\le x\le 65\ {\rm kpc}$, and define the cold front as the $y$-position at which this averaged magnetic field strength attains its maximum. 
In the transverse slice upstream of the reverse shock, regions with $v_y > 3000~{\rm km~s^{-1}}$ are identified as the jet beam, and the jet width $2r_{\rm jet}$ is measured from their transverse extent.
The inner cocoon is defined as the region with the minimum $v_y$ in the same slice, and its pressure is adopted as the inner-cocoon pressure $p_{\rm c}$.
The magnetic energy on the cold front is evaluated using the value of the magnetic energy density, $e_{\rm mag} = (B_x^2 + B_y^2)/8\pi$, averaged over the range $55\ {\rm kpc}\le x\le 65\ {\rm kpc}$ at the $y$-position at the interaction point between the jet and the cold front.
The kinetic energy at the jet head is estimated by calculating the kinetic energy density, $e_{\rm kin} = \frac{1}{2}\rho (v_x^2 + v_y^2)$, in the region defined as the jet head. 

Varying the conduction efficiency parameter $f$ leads to systematic changes in the jet structure. 
As $f$ increases, the pressure within the inner cocoon increases following a scaling relation of $p_{\rm c} \propto f^{0.21}$, while the jet-beam radius decreases as $r_{\rm jet} \propto f^{-0.22}$~(figure~\ref{fig:f_rjet_Pc}).
These trends indicate that more efficient thermal conduction enhances the pressure support of the inner cocoon and results in a more tightly collimated jet.
The monotonic increase of cocoon pressure with $f$ may also reflect a feedback process, in which stronger conduction enhances jet collimation, leading to higher jet-head pressure and further heat transport toward the cocoon. 

\begin{figure}[htbp]
    \centering
    \includegraphics[width=\linewidth]{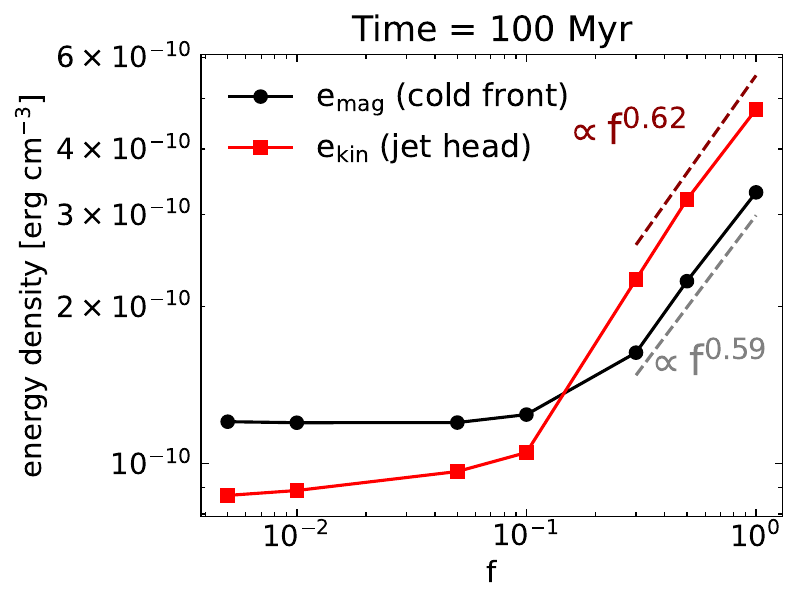}
    \caption{Maximum magnetic energy on the cold-front surface (black) and kinetic energy at the jet head (red) at $t=100\,\mathrm{Myr}$, shown as functions of the thermal conduction efficiency parameters $f$.
    {Alt text: Relationship between thermal conduction efficiency and energy densities at one hundred megayears. The plot shows maximum magnetic energy density on the cold front and kinetic energy density at the jet head in ergs per cubic centimeter as functions of the thermal conduction efficiency parameter. Results are shown for multiple model values of the efficiency parameter.}}
    \label{fig:emag_ekin_vs_f}
\end{figure}

Figure~\ref{fig:emag_ekin_vs_f} shows the dependence of the magnetic energy density on the cold-front surface, $e_{\rm mag}$, and the kinetic energy density at the jet head, $e_{\rm kin}$, on the thermal conduction efficiency parameter $f$.
For $f \le 0.1$, the magnetic energy on the cold-front surface shows little variation with $f$.
In contrast, for $f > 0.1$, the magnetic energy increases systematically with increasing thermal conduction efficiency parameter, following a scaling relation $e_{\rm mag} \propto f^{0.59}$.
This scaling is consistent with the dependence of the kinetic energy at the jet head, which scales as $e_{\rm kin} \propto f^{0.66}$.
The close agreement between these two scalings indicates that the enhanced kinetic energy at the jet head, resulting from stronger thermal conduction, leads to a stronger compression and stretching of magnetic field lines at the cold-front interface, thereby amplifying the magnetic energy.
Overall, these results suggest that thermal conduction influences the magnetic field amplification on the cold-front surface primarily through its impact on the jet dynamics, rather than through direct thermal effects alone.

%
%
\section{Discussion}

\subsection{Conductive Collimation Mechanism of AGN Jets}
\label{sec:dis_CCmechanism}

The simulations indicate systematic differences between the models w/o~TC and w/~TC (subsection~\ref{sec:result_jet}). 
In the w/~TC case, the pressure inside the inner cocoon $p_{\rm c}$ is higher, leading to a narrower jet width $2r_{\rm jet}$ and hence a more collimated jet. 
As the jet becomes more collimated, its momentum along the propagation direction increases, which in turn enhances the magnetic field strength on the cold-front surface (subsection~\ref{sec:result_mag}).
Furthermore, this effect becomes monotonically with increasing conduction efficiency parameter $f$ (subsection~\ref{sec:result_f}).

This behavior can be interpreted in terms of a \textit{conductive collimation mechanism} (hereafter CCM), a framework introduced in this work to describe the role of thermal conduction in shaping the jet structure (figure~\ref{fig:schematic_comparison_of_jet}). 
As the jet interacts with the ICM, the beam is decelerated at the reverse shock, forming a hot region at the jet head. 
Because the magnetic field is predominantly aligned with the jet axis, thermal conduction transports this heat backward along the beam, rather than allowing it to accumulate locally at the jet head. 
This redistribution of thermal energy raises the pressure within the inner cocoon, producing an inward pressure gradient that acts to compress and collimate the jet beam. 

\begin{figure}
    \centering
    \includegraphics[width=\linewidth]{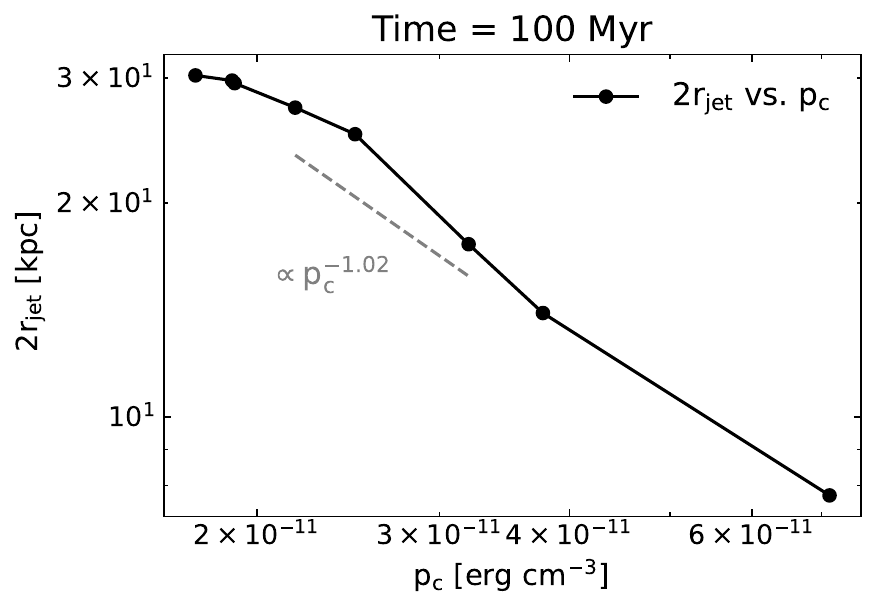}
    \caption{Relation between the inner cocoon pressure and the jet beam width for models with different thermal conduction efficiency parameters $f$.
    {Alt text: Relationship between inner cocoon pressure and jet beam width at one hundred megayears. The plot shows jet beam width in kiloparsecs as a function of inner cocoon pressure in ergs per cubic centimeter for models with different thermal conduction efficiency parameters.}}
    \label{fig:Pc_rjet}
\end{figure}

The jet beam width is regulated by the pressure balance between the jet and the inner cocoon. Denoting the jet beam radius by $r_{\rm jet}$, the jet beam pressure by $p_{\rm jet}$, the jet luminosity by $L_{\rm jet}$, and the inner cocoon pressure by $p_{\rm c}$, the transverse size of the jet can be estimated by assuming pressure equilibrium at the jet boundary, $p_{\rm jet}\simeq p_{\rm c}$.
If the jet beam pressure is determined by the jet luminosity as $p_{\rm jet}\sim L_{\rm jet}/(2 r_{\rm jet} d_{\rm jet} v_y)$, the jet radius scales as 
\begin{equation}
    r_{\rm jet} \sim \frac{L_{\rm jet}}{2 d_{\rm jet} v_y p_{\rm c}}\propto p_{\rm c}^{-1},
\end{equation}
where $d_{\rm jet}$ represents the jet thickness along the line-of-sight (depth direction) and is assumed to be constant.
This simple pressure-balance argument predicts that a higher inner cocoon pressure leads to a narrower, more collimated jet.
Figure~\ref{fig:Pc_rjet} shows the relation between the measured jet beam width and the inner cocoon pressure for models with different thermal conduction efficiencies. The jet radius decreases monotonically with increasing cocoon pressure, and the best-fit power-law relation is consistent with $r_{\rm jet}\propto p_{\rm c}^{-1}$, indicating that enhanced cocoon pressure plays a key role in regulating jet collimation.

\subsection{Dependence on Initial Magnetic Field Geometry}

\begin{figure}
    \centering
    \includegraphics[width=\linewidth]{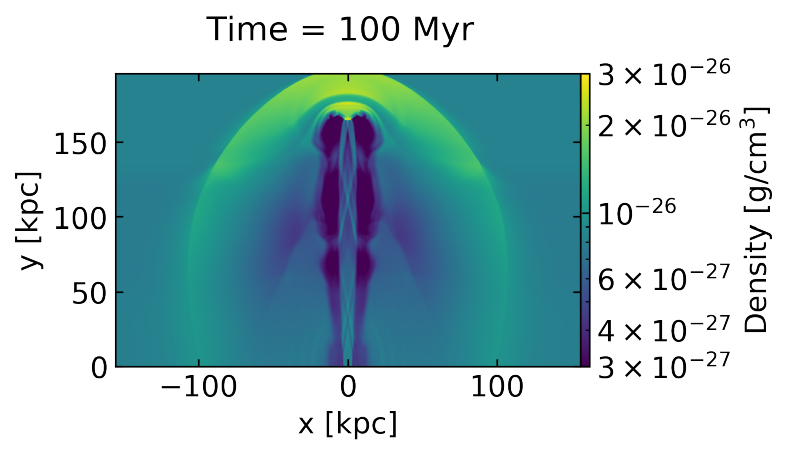}
    \caption{Two-dimensional density plot at $t = 100~\mathrm{Myr}$ for model with plain magnetic fields. 
    {Alt text: Two dimensional gas density distribution at one hundred megayears in a model with plain magnetic fields. The plot shows gas density in grams per cubic centimeter as a function of position in the x and y directions. Spatial axes are in kiloparsecs.}}
    \label{fig:sheet_field_result}
\end{figure}

Since thermal conduction occurs only along magnetic field lines, the initial configuration of the magnetic field possibly affects the outcome of the simulations. To investigate this dependency, we considered a simplified initial magnetic field structure: a uniform sheet field perpendicular to the jet axis~(figure~\ref{fig:sheet_field_result}).

\begin{figure}
    \centering
    \includegraphics[width=\linewidth]{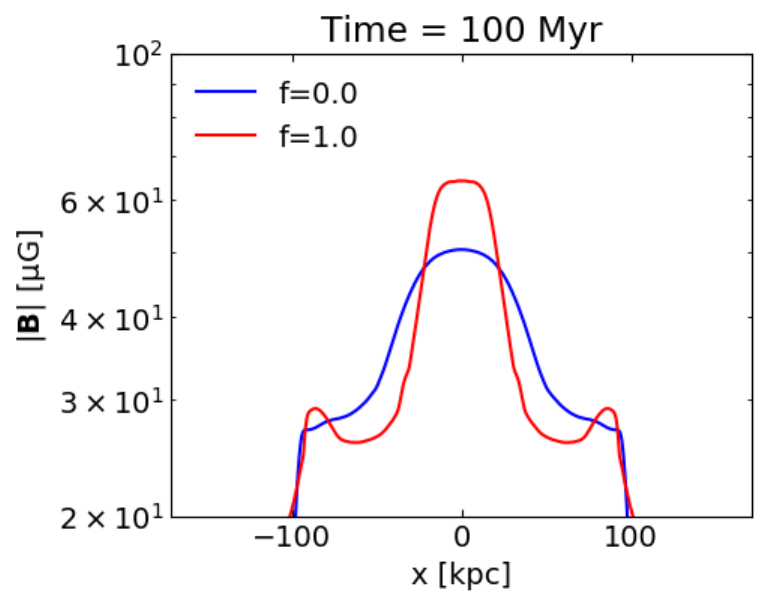}
    \caption{One-dimensional profiles of the magnetic field strength at $t=100~\mathrm{Myr}$, obtained by extracting the maximum value along the $y$-direction at each $x$-coordinate.
    {Alt text: One dimensional profiles of magnetic field strength at one hundred megayears as a function of distance along the x direction. The vertical axis shows magnetic field strength in microgauss and the horizontal axis shows distance in kiloparsecs. Profiles are shown for models without and with thermal conduction.}}
    \label{fig:1d_mag_plain}
\end{figure}

Similar to the case with arch-like magnetic fields, thermal conduction collimates the jet and enhances the magnetic field strength near the interaction region between the jet and the cold front~(figure~\ref{fig:1d_mag_plain}). However, the resulting jet morphology remains axisymmetric because thermal conduction occurs strictly along the $x$-aligned magnetic field lines.

\subsection{Relation to Previous Studies}
Previous studies have suggested that magnetic draping at cold-front surfaces may play an important role in the interaction between AGN jets and the intracluster medium (ICM) (e.g., Asai et al. 2004, 2005, 2007; Markevitch \& Vikhlinin 2007). 
In particular, C21 interpreted the morphology of the radio galaxy MRC~0600--399 in Abell~3376 as a jet being bent by magnetic fields aligned along the cold-front surface.
In the present study, we investigated the jet--ICM interaction using two-dimensional MHD simulations including thermal conduction effects. 
In contrast to the scenario proposed by C21, we do not find strong jet bending caused by the ambient magnetic field. 
This difference may be partly attributed to the two-dimensional nature of our simulations, in which intrinsically three-dimensional processes such as numerical magnetic reconnection are not fully captured. 
In three-dimensional systems, numerical magnetic reconnection and turbulence-driven magnetic-field reorganization may modify the magnetic structure around the jet and potentially influence the jet trajectory. 
Further investigations using fully three-dimensional MHD simulations will therefore be required to examine this possibility.

On the other hand, we find that anisotropic thermal conduction along magnetic field lines accumulated near the contact discontinuity of the jet redistributes thermal energy within the jet. 
This process enhances the collimation of the jet, which we refer to as the \textit{conductive collimation mechanism} (CCM). 
This result does not contradict previous interpretations in which magnetic draping influences the jet morphology, such as \citet{Lyutikov06}; rather, it suggests a complementary process in which heat transport inside the jet modifies its structure and thereby affects the jet--ICM interaction.
Furthermore, our simulations show that the CCM-induced change in jet morphology can lead to the accumulation of magnetic fields near the cold-front interface, and that the resulting magnetic-field strength and distribution may depend on the jet morphology.
While C21 mainly focused on how the magnetic-field structure at the cold front affects jet propagation, our results suggest the complementary possibility that physical processes inside the jet can alter the jet morphology and consequently influence the surrounding magnetic-field structure in the ICM.

Thermal conduction in the ICM has traditionally been discussed mainly in the context of regulating the thermal structure of galaxy clusters (\cite{Sarazin88}; \cite{Narayan01}; \cite{Voit05}). 
Our results suggest that anisotropic thermal conduction may also influence the internal structure and collimation of AGN jets, indicating that heat transport may play an important role in the jet--ICM interaction.
In addition, numerical studies on accretion-disk scales have reported that anisotropic thermal conduction can enhance jet collimation and acceleration (Rezgui et al. 2022, 2025). 
Our results are qualitatively consistent with these studies and suggest that thermal conduction may influence jet structures even on galaxy-cluster scales.

\subsection{Limitations}
In this study, we performed two-dimensional MHD simulations to investigate how anisotropic thermal conduction along magnetic fields affects AGN jets and their interaction with the cold front in the ICM. The CCM proposed in this work arises from magnetic draping within the jet, which leads to the accumulation of magnetic fields near the contact discontinuity. Thermal conduction along these draped magnetic fields redistributes thermal energy within the jet toward the inner cocoon and results in a more collimated jet structure. Because the key elements of this mechanism are the accumulation of magnetic fields by magnetic draping and the resulting anisotropic heat transport along the field lines, CCM is not expected to depend strongly on the dimensionality of the system. Magnetic draping has also been confirmed in three-dimensional jet simulations (e.g., \cite{Lyutikov06}). Our results further indicate that the morphological changes of the jet induced by CCM can enhance the magnetic-field strength near the cold front interface. Therefore, the main objective of this study, clarifying the role of thermal conduction in shaping jet morphology and regulating the interaction between the jet and the ICM, can be addressed within the framework of two-dimensional simulations. However, the present model cannot capture fully three-dimensional instabilities or turbulence. In particular, turbulence may tangle magnetic field lines and reduce the effective thermal conductivity. Evaluating these effects on CCM will require future three-dimensional simulations. Such studies will also be important for more direct comparisons with observations of interactions between AGN jets and cold fronts.

We assumed an arch-like magnetic-field configuration in the ICM. Such a configuration can be regarded as a simplified representation of the spiral magnetic structures expected around cold fronts and sloshing regions in merging galaxy clusters~(e.g., \cite{Vikhlinin02}; \cite{Markevitch07}). Previous studies have shown that even when the surrounding ICM is turbulent, magnetic draping can align magnetic fields around moving structures (e.g., Asai et al. 2007; \cite{Dursi08}), which suppresses thermal conduction across the field lines. 
This suggests that CCM may not strongly depend on the detailed global magnetic-field configuration. 
Indeed, we confirmed that CCM also operates in simulations with a uniform sheet field perpendicular to the jet axis. 
However, the magnetic-field configuration still affects the jet morphology; for example, the jet tends to remain more axisymmetric in the uniform sheet field case than in the arch-like configuration. 
This indicates that the magnetic-field geometry may influence the symmetry and bending of the jet. 
Exploring the properties of CCM under more realistic ICM magnetic-field structures will therefore be an important subject for future work.

Additional physical processes are not included in the present study. 
Viscosity and plasma instabilities may reshape the jet and magnetic fields on the cold-front surface, and cosmic-ray populations and their associated synchrotron emission are not modeled in detail. 
Since radio observations provide key diagnostics of jet bending and cold front evolution, incorporating cosmic-ray transport and synchrotron cooling will be essential for direct comparison with observed radio morphologies.
Finally, plasma associated with the central galaxy is neglected, even though such material can extend to radii of tens of kpc and may modify the jet environment. 
While the adopted jet parameters fall within the observationally inferred range for AGN-driven outflows, uncertainties in jet energetics, magnetization, and composition remain inherent to the problem. 
Future work will benefit from extending the simulations to fully three–dimensional geometries, including additional plasma processes and cosmic-ray physics, in order to evaluate the robustness and observational consequences of the conductive collimation mechanism.

\section{Conclusion}
This study has examined the influence of anisotropic thermal conduction on the evolution of AGN jets in a galaxy cluster environment using two-dimensional MHD simulations. 
The simulations indicate that thermal conduction can redistribute heat from the jet head into the inner cocoon along magnetic field lines, modifying the pressure balance that governs the interaction between the jet beam and its surroundings. 
This redistribution acts to maintain a more collimated jet structure, which enhances the interaction with the ambient intra-cluster magnetic field and leads to localized magnetic field compression and amplification near the jet–ICM interface.

These results may be interpreted within a Conductive Collimation Mechanism~(CCM), in which thermal transport along ordered magnetic fields contributes to the degree of jet collimation and the associated magnetic field amplification. 
In this framework, thermal conduction is not simply a diffusive process, but participates in shaping the morphology and field structure of the jet–cocoon system. 
The findings suggest that incorporating thermal conduction is relevant when evaluating the structure and strength of magnetic fields in merging cluster environments.
Overall, the results highlight the value of considering multiple plasma processes when modeling AGN jet interactions with the ICM. 
Future work including fully three-dimensional magnetic field geometries, cosmic-ray transport, and direct comparison with radio and X-ray observations will help further assess the applicability and observational consequences of the CCM in cluster systems.

\begin{ack}
Numerical computations were in part carried out on the PC cluster at the Yukawa Institute for Theoretical Physics (YITP), Kyoto University, and at the Center for Computational Astrophysics (CfCA), National Astronomical Observatory of Japan.
This research was supported by the International Research Exchange Support Program of the National Institutes of Natural Sciences.
This work was supported in part by JSPS KAKENHI (TA:21H04492, MM:22H01272, 23K22543).

\end{ack}

\section*{Supplementary data}
The following supplementary data is available in the online version of this article.
E-movies 1--2.

\end{document}